\documentclass[twocolumn,twocolappendix,times,astrosymb,tighten]{aastex701}

\newcommand{\wtp}{WTP14adeqka}
\newcommand{\aips}{$\mathcal{AIPS}$}

\newcommand{\hms}[4]{\ensuremath{{#1}^{\rm h}{#2}^{\rm m}{#3}\fs{#4}}}
\newcommand{\dms}[4]{\ensuremath{{#1}\degr{#2}\arcmin{#3}\farcs{#4}}}

\received{August 22, 2025}

\submitjournal{ApJ}

\shorttitle{\wtp: First Resolved Delayed Outflow}
\shortauthors{Golay et al.}

\begin{document}

\title{Radio Emission from the Infrared Tidal Disruption Event WTP14adeqka: The First Directly Resolved Delayed Outflow from a TDE}

\author[0000-0001-7946-1034]{Walter W. Golay}
\affiliation{Center for Astrophysics $|$ Harvard \& Smithsonian, 60 Garden St., Cambridge, MA 02138, USA}
\email[show]{wgolay@cfa.harvard.edu}
\correspondingauthor{Walter W. Golay}

\author[0000-0002-9392-9681]{Edo Berger}
\affiliation{Center for Astrophysics $|$ Harvard \& Smithsonian, 60 Garden St., Cambridge, MA 02138, USA}
\email{eberger@cfa.harvard.edu}

\author[0000-0001-7007-6295]{Yvette Cendes}
\affiliation{Department of Physics, University of Oregon, 1030 E 13th Ave., Eugene, OR 97403, USA}
\email{yncendes@uoregon.edu}

\author[0000-0003-4127-0739]{Megan Masterson}
\affiliation{MIT Kavli Institute for Astrophysics and Space Research, Massachusetts Institute of Technology, Cambridge, MA 02139, USA}
\email{mmasters@space.mit.edu}

\author[0000-0003-3272-9237]{Emil Polisensky}
\affiliation{U.S. Naval Research Laboratory, 4555 Overlook Ave SW, Washington, DC 20375, USA}
\email{emil.j.polisensky.civ@us.navy.mil}

\author[0000-0003-1511-6279]{Robert L. Mutel}
\affiliation{Department of Physics \& Astronomy, University of Iowa, 30 N Dubuque St., Iowa City, IA 52242, USA}
\email{robert-mutel@uiowa.edu}

\author[0000-0003-0526-2248]{Peter K. Blanchard}
\affiliation{Center for Astrophysics $|$ Harvard \& Smithsonian, 60 Garden St., Cambridge, MA 02138, USA}
\email{peter.blanchard@northwestern.edu}

\author[0000-0003-0871-4641]{Harsh Kumar}
\affiliation{Center for Astrophysics $|$ Harvard \& Smithsonian, 60 Garden St., Cambridge, MA 02138, USA}
\email{harsh.kumar@cfa.harvard.edu}

\author[0000-0003-4768-7586]{Raffaella Margutti}
\affiliation{Department of Astronomy, University of California, Berkeley, CA 94720-3411, USA}
\email{rmargutti@berkeley.edu}

\author[0000-0001-7081-0082]{Maria Drout}
\affiliation{David A. Dunlap Department of Astronomy and Astrophysics, University of Toronto 50 St. George Street, Toronto, Ontario, M5S 3H4, Canada}
\email{maria.drout@utoronto.ca}

\author[0009-0001-9034-6261]{Christos Panagiotou}
\affiliation{MIT Kavli Institute for Astrophysics and Space Research, Massachusetts Institute of Technology, Cambridge, MA 02139, USA}
\email{cpanag@mit.edu}

\author[0000-0002-8989-0542]{Kishalay De}
\affiliation{Department of Astronomy and Columbia Astrophysics Laboratory, Columbia University, 550 W 120th St. MC 5246, New York, NY 10027, USA}
\affiliation{Center for Computational Astrophysics, Flatiron Institute, 162 5th Ave., New York, NY 10010, USA}
\email{kd3038@columbia.edu}

\author[0000-0003-0172-0854]{Erin Kara}
\affiliation{MIT Kavli Institute for Astrophysics and Space Research, Massachusetts Institute of Technology, Cambridge, MA 02139, USA}
\email{ekera@mit.edu}

\begin{abstract}
We present detailed radio observations of the mid-infrared (MIR) tidal disruption event (TDE) \wtp.  We detect rising radio emission starting $\approx  4$ years after the discovery of the MIR emission (and about 2 years after its peak), peaking at $\approx 6.5$ years and declining thereafter, reminiscent of the delayed radio emission recently identified in optically discovered TDEs. The peak radio luminosity, $\nu L_\nu\approx 2\times 10^{39}$ erg s$^{-1}$, is comparable to the brightest radio emission in optical TDEs. Multi-frequency radio observations at 8.9 and 9.7 years reveal a non-relativistic outflow with a mean expansion velocity of $\approx 0.021c$ (for an assumed launch at the time of disruption) and an energy of $\approx 10^{50.7}$ erg, about an order of magnitude larger than in typical optical TDEs. More importantly, Very Long Baseline Array (VLBA) observations at the same epochs directly resolve the radio source and reveal an increase in size from approximately 0.11 pc to 0.13 pc (with no apparent astrometric shift), corresponding to an expansion velocity of $\approx 0.05c$, and a likely delayed launch by about 2 years. The VLBA size measurements rule out an off-axis jet launched at the time of disruption, which would have an expected size of $\gtrsim {\rm pc}$ on these timescales; the possibility of a delayed jet can be evaluated with future VLBA observations. We conclude that MIR TDEs can launch energetic, delayed outflows. Ongoing radio observations of the full MIR TDE sample will reveal whether this behavior is ubiquitous.
\end{abstract}


\section{Introduction 
\label{sec:intro}}

The tidal disruption of a stray star by a supermassive black hole (SMBH) powers transient accretion and energetic ejection of stellar debris \citep{hills_possible_1975, rees_tidal_1988}, which manifests across the electromagnetic spectrum as a tidal disruption event (TDE). Radio observations of synchrotron emission from TDE ejecta interacting with the circumnuclear medium allow us to infer the energetics and geometry of the outflow \citep{alexander_radio_2020}, which can reveal the presence and properties of non-relativistic quasi-spherical outflows, on-axis jets \citep[e.g.,][]{berger_radio_2012,cendes_radio_2021}, or off-axis jets \citep[e.g.,][]{matsumoto_generalized_2023}. Studying transient outflows from SMBHs, such as those triggered by tidal disruption events (TDEs), may reveal how these outflows depend on the underlying SMBHs' properties \citep[e.g., spin \& mass,][]{van_velzen_seventeen_2021}.

Recent observations by \citet{cendes_ubiquitous_2024} indicate that $\approx\frac{1}{3}$ of optically-discovered TDEs exhibit radio emission delayed by $\sim 1000$\,days. This time lag may be due to off-axis jets \citep[e.g.,][]{cendes_mildly_2022, matsumoto_generalized_2023, beniamini_swift_2023}, a previously unknown phase of the outflow \citep[e.g.,][]{hajela_eight_2024}, a complex density structure surrounding the SMBH \citep{matsumoto_late-time_2024}, or delayed, non-relativistic outflows \citep[e.g.,][]{horesh_delayed_2021, horesh_are_2021}. In a systematically-selected sample of X-ray-selected TDEs discovered by the eROSITA telescope, 8 (36\%) of candidates displayed radio variability and 6 (27\%) had variability exceeding that expected from interstellar scintillation at $>500$\,d post-discovery. \citet{goodwin_systematic_2025} found no significant difference between the X-ray-selected sample and the optical sample of \citet{cendes_ubiquitous_2024} in the detection rate, the radio luminosities, or some outflow properties like minimum outflow radius and velocity. Tentatively, outflow energies may be slightly higher in the X-ray sample. However, the optical sample was more likely to still be rising in the radio at $>1000$\, post-discovery. In contrast, all eight of the radio-detected X-ray TDEs began fading by this timescale.

The exploration of delayed radio emission has thus far been limited to optically and X-ray discovered TDEs \citep{cendes_ubiquitous_2024, goodwin_systematic_2025}. Some optical and X-ray TDEs have shown evidence of MIR dust echoes \citep{dou_long_2016, jiang_wise_2016, van_velzen_discovery_2016}; however, the relative dust-covering factors are low \citep[$\sim$few\%,][]{jiang_wise_2016, jiang_infrared_2021}. In the case of large dust-covering fractions, the initial optical/UV/X-ray emission that is reprocessed by circumnuclear dust \citep{van_velzen_reverberation_2021} to the IR may be the only way to detect the TDE. Such a population could explain the apparent lack of TDEs in dusty star-forming galaxies as an observational bias \citep{french_tidal_2016}. 

Recently, \citet{panagiotou_luminous_2023} and \citet{masterson_new_2024} used archival WISE data to identify a sample of 12 MIR TDEs, none of which exhibit any evidence of past AGN activity, and most have no associated optical transient. This sample doubled the number of TDEs known within 200\,Mpc, suggesting that roughly half of all TDEs may be dust-obscured. The MIR TDEs span a range of inferred disruption dates from 2014 to 2018, thus providing a compelling sample for the search for delayed radio emission as seen in optical and X-ray TDEs of comparable ages \citep{cendes_ubiquitous_2024, goodwin_systematic_2025}. 

Here we focus on the MIR TDE \wtp\ (angular distance $d_A=79.3\,{\rm Mpc}$), which was first detected by WISE on 2015 June 20 and exhibited one of the most luminous MIR light curves in the sample, and had no associated detections in optical or X-ray \citep{masterson_new_2024}. The flare peaked about two years after the initial detection with a luminosity of $L_{\rm MIR}\approx 10^{43}$\,erg\,sec$^{-1}$ and was detected through the remainder of the WISE mission, which de-orbited in 2024. Modeling of the MIR light curves indicated reprocessing of the TDE light by a spherical dust shell with a radius of $0.33^{+0.13}_{-0.05}$~pc, and a model of the host galaxy SED to extract a star-forming mass predicts a central SMBH mass of $\log M_{\rm SMBH}\approx6.8\pm0.5$ \citep{masterson_new_2024}.

In this paper, we present a detailed radio study of \wtp\ using archival and targeted observations spanning $\approx 4-10$ years after disruption, making this event the first MIR TDE with detected delayed radio emission. The observations include data at multiple frequencies and high-resolution imaging with very long baseline interferometry (VLBI). We use an equipartition analysis and directly resolved imaging of the radio source to determine the outflow properties. The paper is structured as follows. In \S\ref{sec:obs} we present the radio observations and optical spectroscopy of the host galaxy. In \S\ref{sec:host} we discuss the host galaxy properties and show that pre-existing radio emission is due to star formation, ruling out any contribution from an AGN. In \S\ref{sec:outflow} we determine the outflow and circumnuclear properties from an equipartition analysis of multi-frequency radio observations, as well as from direct measurements of the source size and expansion rate from VLBI data, and explore the implications for the outflow geometry and launch timescale.  We compare the outflow properties of \wtp\ to those of optical TDEs in \S\ref{sec:discussion}, and conclude and summarize in \S\ref{sec:summary}. Throughout the paper we use cosmological parameters $\Omega_M=0.3$, $\Omega_\Lambda=0.7$, and $H_0=70$\,km\,s$^{-1}$\,Mpc$^{-1}$. 

\begin{figure*}[ht]
    \centering
    \includegraphics[width=\linewidth]{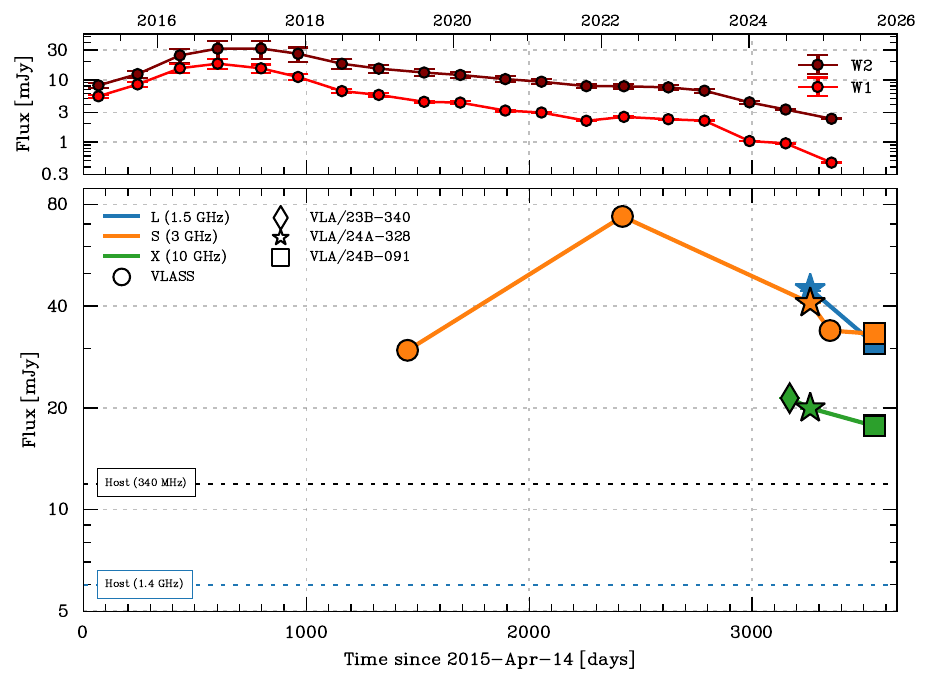}
    \caption{Radio light curves of \wtp\ in selected frequencies from archival and targeted VLA observations (main panel) along with the MIR TDE light curve \citep[where we have updated the light curve from][with additional data through the end of the WISE mission and refined errors]{masterson_new_2024}. We show radio data at frequencies that have $\ge 3$ data points, including the pre-transient NVSS 1.4\,GHz detection (blue dashed line), all three VLASS 3\,GHz epochs (orange circles), our targeted programs, and commensal VLITE 340\,MHz detections (black dashed line) that we attribute to the host galaxy (\S\ref{sec:host}). The ``zero time'' is set by a best-fit model to the MIR light curve of 2015 April 14 \citep[see][for further details]{masterson_new_2024}.}
    \label{fig:radio-lc}
\end{figure*}

\section{Observations 
\label{sec:obs}}

The MIR TDE \wtp\ ($z=0.01895\pm0.0001$; luminosity distance $d_L=82.3$\,Mpc; $d_A=79.3$\,Mpc; 2.60\,mas\,pc$^{-1}$, \citealt{strauss_redshift_1992}) was first detected in the MIR on 2015 June 20 at $\alpha=\hms{19}{49}{24}{86},\ \delta=\dms{+63}{30}{33}{4}$ (J2000). Modeling of the light curve as a single-radius dust echo indicates a disruption date of 2015 April 14 \citep{masterson_new_2024}, which we adopt as the reference epoch for all dates reported here.

\subsection{Very Large Array 
\label{subsec:vla}}

We performed targeted multi-frequency radio observations using the Karl~G.~Jansky Very Large Array (VLA) spanning the entire frequency range of $\approx 1-50$ GHz, and additionally compiled archival data from past and ongoing surveys; see Table~\ref{tab:fluxes}. The first available VLA observation after the MIR detection of \wtp\ is from 2019 April 12 ($+1454$\,d), taken as part of the VLA Sky Survey \citep[VLASS,][]{lacy_karl_2020} in S-band ($2-4$\,GHz); two additional VLASS observations are available, from 2021 December 1 ($+2418$\,d) and 2024 June 19 ($+3349$\,d). These observations reveal a factor of $\approx 3$ brightening and dimming in the source brightness. Following the publication of \wtp\ \citep{masterson_new_2024}, and recognizing its radio detections in the first two epochs of VLASS, we obtained dedicated observations on 2023 December 21 ($+3168$\,d) in X-band ($8-12$\,GHz; Project ID: VLA/23B-340, PI: Golay), as well as on 2024 March 22 ($+3260$\,d, Project ID: VLA/24A-328, PI: Cendes) and 2025 January 3 ($+3547$\,d, Project ID: VLA/24B-091, PI: Golay) in all VLA frequencies. We used the standard NRAO VLA Pipeline (v2024.1.1.22), included in the Common Astronomy Software Applications \citep[CASA v6.6.1,][]{casa_team_casa_2022} to calibrate and analyze the dedicated observations. We accessed the VLASS observations using the ``Quicklook'' data products available via the Canadian Astronomy Data Centre\footnote{\href{https://www.cadc-ccda.hia-iha.nrc-cnrc.gc.ca/en/search/?collection=VLASS&noexec=true}{https://www.cadc-ccda.hia-iha.nrc-cnrc.gc.ca/en/search/?collection=VLASS}}. The radio light curves at all frequencies that have at least three data points are shown in \autoref{fig:radio-lc} (alongside the MIR light curve for comparison).

During the above observations, the VLA Low-band Ionosphere and Transient Experiment \citep[VLITE,][]{clarke_commensal_2016, polisensky_exploring_2016} observed the field commensally at P-band ($320-360$\,MHz). A search of the VLITE archive revealed additional observations on 2024 November 9 ($+3492$\,d), centered $0.96^\circ$ from the position of \wtp\ (Project ID: VLA/24B-448, PI: Dong). A source is consistently detected in most VLITE epochs. We combined all epochs from the VLITE Commensal Sky Survey (VCSS) into a single dataset, which yielded a flux density of $11.9\pm 2.3$\,mJy (see \autoref{fig:radio-lc}). Due to the changing VLA configurations, the VLITE synthesized beam ranged from $\sim 5-45^{\prime\prime}$ across the various epochs. No correlation is observed between the measured flux density and resolution, suggesting that the source is unresolved. However, we caution that since the detections are of modest significance (${\rm SNR}\approx 4-6$), the relatively large uncertainties may obscure subtler resolution-dependent effects.

The VLITE measurements are consistent with a steady radio source at the position of \wtp, which we attribute to emission from the host galaxy, UGC~11487. This interpretation is further supported by the detection of a coincident source in the NRAO VLA Sky Survey \citep[NVSS,][]{condon_radio_2002}, which observed the field on 1995 April 2 ($\approx 20$\,years before \wtp). The NVSS source exhibits a flux density of $6.0\pm 0.5$\,mJy at L-band (1.4\,GHz; see \autoref{fig:radio-lc}). A more detailed discussion of the host galaxy emission is presented in \autoref{sec:host}.

\subsection{Very Long Baseline Array \label{subsec:vlbi}}

Following the VLA detections, and taking advantage of the brightness and proximity of \wtp, we obtained high-resolution observations with the NRAO Very Long Baseline Array (VLBA) on 2024 January 2 ($+3180$\,d, Program ID: BG289A, PI: Golay) and 2025 January 25 ($+3569$\,d, Program ID: BM569A, PI: R. Margutti \& M. Drout). The receivers were centered at 8.368\,GHz and used the dual Roach Digital Backend (RDBE2) setup in the Digital Downconverter (DDC) mode. The four intermediate frequency windows (IFs) span 128\,MHz each in dual polarization (R+L) and are recorded by the {\tt Mark\,VI} system at 2\,bits for an aggregate baseband recording rate of 4\,Gbps. This mode provides maximal bandwidth and sensitivity at the expense of less accurate absolute flux calibration; however, this is not a significant concern since our VLA observations provide accurate ($\approx 7\%$) flux density calibration.

We used the nodding phase-referencing observation scheme \citep{lestrade_phase-reference_1990}, scheduling station pointings that cycled between the primary phase calibrator J2006+6424 and \wtp\ with a 3-min duty cycle consisting of 30\,s dwelling on the calibrator followed by a 2.5-min pointing of \wtp. Additionally, every $\approx 20$\,min, we phase-referenced to two nearby calibrators, J1959+6206 and J1938+6307, that enclose the position of \wtp\ as a group for MultiView \citep[e.g.,][]{reid_microarcsecond_2014}. We also started and ended the 8-hr observations with 30-min geodetic blocks, as well as a 20-min geodetic block in the middle of the observation. These secondary checks throughout the observations enable second-order calibration corrections for maximizing astrometric precision \citep{pradel_astrometric_2006,reid_microarcsecond_2014}, a requirement for measuring or constraining the proper motion of the radio source \citep{reid_techniques_2022}. The resulting total on-source time is 3.8 hr in both epochs.

\begin{figure*}[ht]
    \centering
    \includegraphics[width=\linewidth]{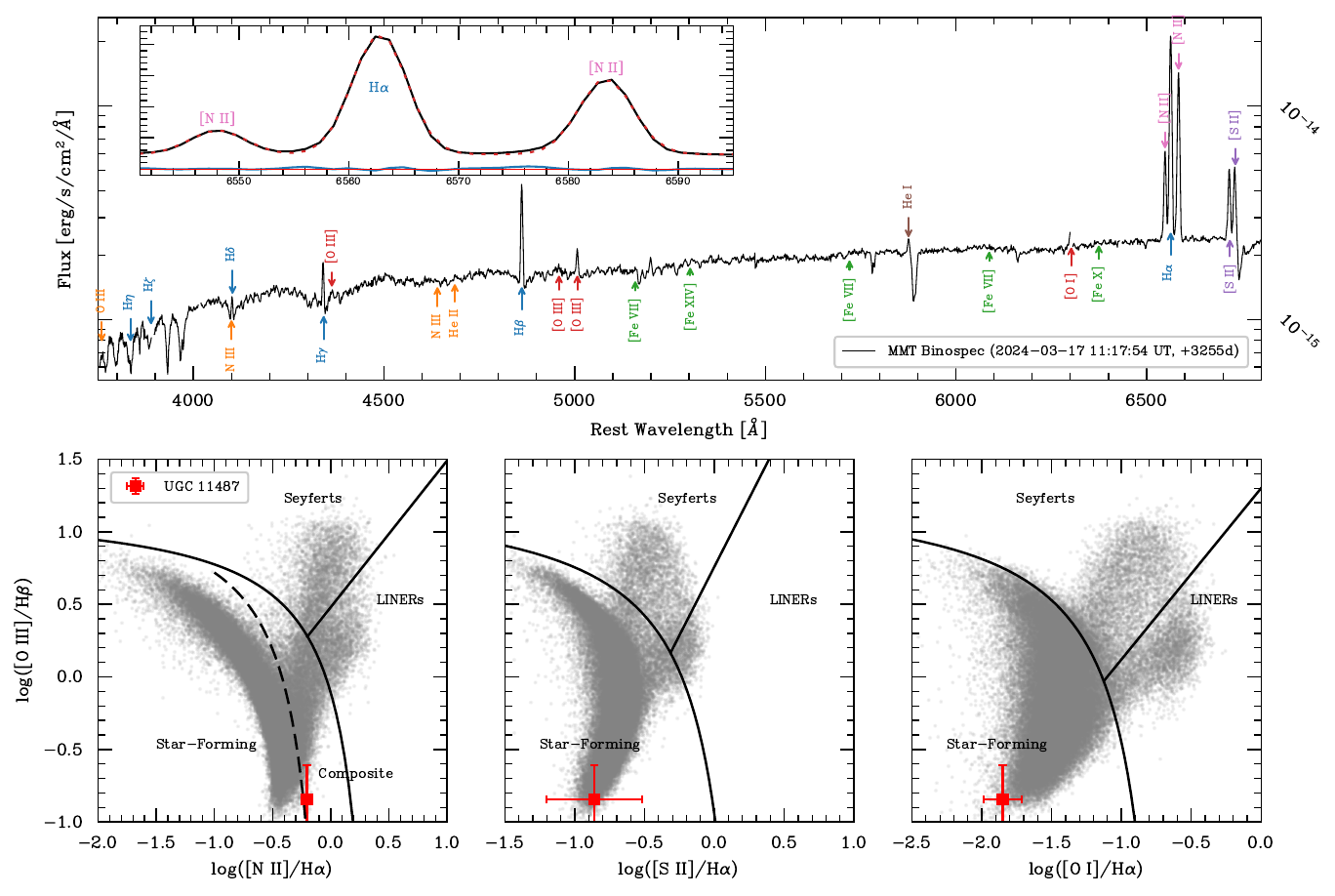}
    \caption{{\it Top:} MMT/Binospec optical spectrum of the central region of the host galaxy of \wtp\ (UGC~11487), with key detected emission lines labeled; also marked are (undetected) Bowen fluorescence lines (orange) and coronal lines (green) that could indicate AGN activity. The inset shows a zoom-in of the H$\alpha$+[\ion{N}{2}] lines, along with a simple three-component Gaussian fit, indicating unresolved lines at $\approx 120$ km s$^{-1}$. {\it Bottom:} Emission line ratio diagnostic plots marking the boundaries used to separate star-formation and AGN-dominated spectra  \citep{kewley_host_2006, wang_spectral_2018}, and showing the distribution for SDSS galaxies at $z<0.1$ \citep[DR18,][]{almeida_eighteenth_2023}. The line ratios for UGC~11487 place it in the star-forming region of the diagrams.}
    \label{fig:optical-spectrum}
\end{figure*}

The phase tracking center was set to the MIR position of \wtp: $\alpha=\hms{19}{49}{24}{86},\ \delta=\dms{+63}{30}{33}{4}$ (J2000; \citealt{masterson_new_2024}). The data were correlated using the DiFX software correlator operated by NRAO in Socorro, NM \citep{deller_difx-2_2011}. The resulting visibilities consist of 256 digitized channels spanning each of the four 128\,MHz IFs with dual polarization. 

We calibrated the VLBA observations using the NRAO Astronomical Image Processing Software \citep[\aips,][]{greisen_aips_2003}. We removed all visibilities observed at an elevation of $<15\degr$ since astrometric uncertainty degrades rapidly at low elevation \citep{reid_microarcsecond_2014}. We then applied standard VLBI delay and amplitude corrections using the {\tt VLBAUTIL} tools in \aips\ \citep{ulvestad_aips_2001} and the geodetic corrections from the three blocks using the \aips\ task {\tt DELZN} \citep{mioduszewski_aips_2009}. Finally, we used the primary and the two secondary calibrators to apply a time-varying atmospheric phase gradient solution to the target using the \aips\ task {\tt ATMCA} \citep{fomalont_aips_2005}. The final calibrated datasets were averaged over all channels for a single broadband measurement in each of the two epochs.

We used the difference mapping program {\tt difmap} \citep{shepherd_difmap_1997} to perform additional flagging, analyze the visibilities, and produce images. To measure the position and size of the radio source, we used the {\tt modelfit} routine to fit an image-plane model directly to the visibilities. This approach offers the benefit of a robust characterization of the uncertainties in the model parameters, as the data variance is not propagated through the ill-posed inverse imaging problem. Tracking uncertainty through procedures like deconvolution \citep[e.g., {\tt CLEAN},][]{hogbom_aperture_1974} is challenging in this context. Visual inspection of the visibility amplitude versus projected $(u,v)$ baseline length (a {\tt projplot} in {\tt difmap}) across a variety of angles indicates a highly resolved symmetric source that is effectively Gaussian in shape. This fact motivates our choice of a single Gaussian source as a simple model for the image plane. Finally, to confirm our characterization of the source, we used the {\tt difmap} hybrid-mapping scripts {\tt automap} and {\tt muppet} \citep{shepherd_difmap_1997} to produce a high-fidelity, low-noise, self-calibrated map of the emission distribution on the sky. 

\subsection{Host Galaxy Optical Spectroscopy}

Given the pre-existing radio detection at 1.4\,GHz, and the steady radio emission at 340\,MHz after the discovery of \wtp, we assess the potential presence of an AGN using an optical spectrum of the host galaxy obtained with Binospec \citep{fabricant_binospec_2019} on the MMT 6.5\,m telescope observed on 2024 March 17 ($+3255$\,d). We obtained two 600\,s exposures with a $1^{\prime\prime}$ wide slit and the 270\,lpmm grating centered at 6500\,\r{A} providing a spectral resolution of $R\approx 2500$. The slit was centered on the VLBA-derived coordinates. The spectra were reduced, extracted, and calibrated using the latest version of the Binospec Pipeline\footnote{\href{https://bitbucket.org/chil_sai/binospec/src/master/}{https://bitbucket.org/chil\_sai/binospec/src/master/}} using a reference observation of the standard star Feige~34. The continuum signal-to-noise ratio is $\gtrsim 50$ across most of the wavelength range. The flux-calibrated spectrum is shown in \autoref{fig:optical-spectrum}. 

\section{Host Galaxy Properties: No Evidence for an AGN 
\label{sec:host}}

We present several lines of evidence that the host galaxy of \wtp, UGC~11487, is star-forming and lacks detectable AGN activity that could be responsible for the pre- or post-TDE radio emission. In \autoref{fig:optical-spectrum} we place the host galaxy on the standard optical emission line ratio diagnostic plots used for galaxy classification \citep{kewley_host_2006, wang_spectral_2018}, and find that it is located in the star-forming region, driven primarily by the weak [\ion{O}{3}] emission lines. Although previous optical spectra already pointed to this result, they were insufficient to resolve the [\ion{N}{2}]+H$\alpha$ complex for the complete diagnostic diagram \citep{masterson_new_2024}. The narrow width of the lines ($\approx120$\,km\,s$^{-1}$), along with the lack of any Bowen fluorescence or coronal lines associated with AGN activity (\autoref{fig:optical-spectrum}), provides further support for a star formation origin. 

\begin{figure*}[t]
    \centering
    \includegraphics[width=\linewidth]{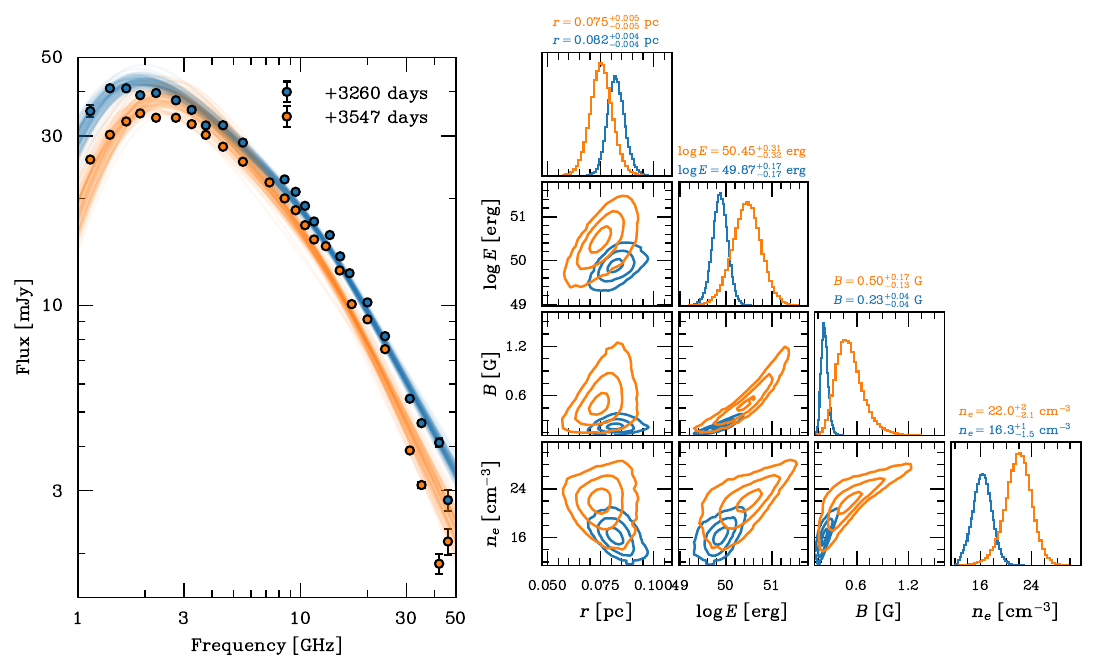}
    \caption{\textit{Left:} Radio spectral energy distributions from 2024 March 22 (blue) and 2025 January 3 (orange), along with 100 model curves from MCMC fitting (\autoref{sec:outflow}). The peak of the SEDs is at $\approx 2$\,GHz, and we further observe a cooling break at higher frequencies, for which we discuss the implications for deviation from equipartition in \autoref{subsec:cooling}. \textit{Right:} Joint posterior model parameter probability distributions (assuming energy equipartition) inferred from the MCMC distributions of the best-fitting peak fluxes and frequencies. Contour levels are shown at 39, 87, and 99 percent, and parameter ranges are reported as the 16th, 50th, and 84th percentiles of the distributions.}
    \label{fig:radio-sed}
\end{figure*}

UGC~11487 was among an early sample of galaxies used to derive the FIR-radio luminosity scaling relationship from NVSS observations combined with \textit{IRAS} data \citep{condon_radio_2002}. This correlation is due to the fact that FIR emission traces thermal dust emission from \ion{H}{2} regions, while the radio emission traces synchrotron radiation from supernova remnants. Using this correlation, with FIR flux densities of $F_{\nu,60\mu {\rm m}}\approx 1.29$\,Jy and $F_{\nu,100\mu {\rm m}}\approx 2.32$\,Jy measured by IRAS \citep{moshir_iras_1990}, we find an expected 1.4\,GHz flux density of $\approx 7$\,mJy, in good agreement with the NVSS value of $\approx 6$\,mJy (\S\ref{subsec:vla}). This result is also supported by the steady emission detected in the VLITE observations at 340\,MHz, at $\approx 12$\,mJy, which indicates a spectral index of $\alpha\approx -0.5$ ($F_\nu\propto \nu^\alpha)$, typical of synchrotron radiation from star-forming regions \citep{tabatabaei_radio_2017}.

We therefore conclude that UGC~11487 displays no evidence for AGN activity in either its optical spectrum or radio emission, and that the pre-existing 1.4\,GHz emission and the post-TDE 340\,MHz emission are both due to star formation in the host galaxy. The lack of an X-ray detection to $L_X\lesssim10^{41.8}$\,erg\,s$^{-1}$ by \textit{Swift XRT} \citep{masterson_new_2024} excludes a majority of X-ray-detected AGN, which are typically $L_X\lesssim10^{42}-10^{46}$\,erg\,s$^{-1}$.

\section{Radio Outflow Properties 
\label{sec:outflow}}

The broadest temporal coverage of \wtp's transient radio emission is in S-band ($2-4$\, GHz), where the radio light curve displays a broad, long-lasting, and delayed brightening and subsequent decline that is observed through our targeted observations at multiple frequencies (\autoref{fig:radio-lc}). The initial radio detection is at $\approx 4$\, years with a flux density of $\approx 30$ mJy, followed by an apparent peak about 2.5 years later at $\approx 74$ mJy. Subsequently, the light curve declines as $F_\nu\propto t^{-2}$. To determine a characteristic behavior of the light curve, we assume a more rapid rise of $F_\nu\propto t^3$ (as observed in some of the optical TDEs with delayed radio emission; \citealt{cendes_ubiquitous_2024}) anchored to the first VLASS detection, would suggest a peak at $\approx 6$ years with a peak flux density of $\approx 100$\,mJy.

While the single-frequency temporal evolution indicates the presence of a transient radio signal, multi-frequency spectral energy distributions (SEDs) are required to characterize the properties of the outflow. The radio SEDs of TDEs are typically well-modeled with a synchrotron self-absorbed (SSA) model, which is expected to be a better fit due to the typically high-density circumnuclear environment; however, free-free absorption often cannot be independently ruled out without broad, low-frequency coverage to distinguish between the optically thick part of the SED. Using the SSA model and assuming equipartition of energy between the radiating electrons and the magnetic fields, we can extract the physical properties of the outflow (e.g., energy, radius).

The two SEDs at 8.9 and 9.7 years are shown in Figure~\ref{fig:radio-sed}. We fit the SEDs with the model of \citet{granot_shape_2002}, originally developed for synchrotron emission from gamma-ray burst (GRB) afterglows, but routinely applied to TDEs (e.g., \citealt{zauderer_illuminating_2013, cendes_radio_2021, cendes_mildly_2022,cendes_ubiquitous_2024}). We use the limit where the SSA frequency is larger than the synchrotron frequency, $\nu_a\gg\nu_m$, relevant for sources that are not ultra-relativistic. Using the procedure of \citet{cendes_ubiquitous_2024}, we fit for four parameters: flux normalization, $F_{\nu}(\nu_m)$, SSA frequency, $\nu_a$, electron power-law energy index, $p$, and synchrotron cooling frequency, $\nu_c$.  We use the Markov Chain Monte Carlo (MCMC) sampler {\tt emcee} \citep{foreman-mackey_emcee_2013}. We also include a nuisance parameter, $\ln f$, as an additional fractional error that quantifies systematic uncertainty as estimated from the data's intrinsic variance. We implement a weak (non-informative) uniform prior on all parameters intended only to constrain the fit to physical values (e.g., non-zero flux and a reasonable upper limit to remain normalizable). We use 100 chains to sample for 10,000 steps. Inspection of the chains revealed rapid convergence; however, we conservatively removed the first 1,000 steps in each chain before proceeding with further analysis.

Our first SED was obtained in the VLA C-configuration, which results in a relatively large synthesized beam leading to some contamination from the spatially extended host galaxy emission; e.g., the L-band (1.5\,GHz) synthesized beam is $\theta_{\rm FWHM}\approx 14^{\prime\prime}$, which is comparable to the size of the host galaxy. To account for the small additional host flux, we use the inferred host radio spectral index of $\approx -0.5$ anchored to the pre-transient NVSS detection of 6 mJy at 1.4 GHz (\S\ref{sec:host}). We subtract the contribution of the host from the L-band and S-band ($3$\,GHz) measurements for which the synthesized beam is large enough to include the extended host, and the images exhibit evidence for some resolved structure. We find a small fractional contribution of about $13\%$ in L-band and $10\%$ in S-band. The best-fitting peak flux density in the first SED is $F_{\nu,p}=42.9\pm 1.1$\,mJy at a peak frequency of $\nu_p=1.95\pm 0.1$\,GHz, with $p=2.31\pm 0.09$, and $\nu_c=12.3^{+1.9}_{-1.6}$\,GHz. The noise term is $\ln f=-2.6\pm0.2$. We attempted the SED fitting procedure without the host galaxy flux correction, and the resulting parameters changed by $\lesssim 10\%$. The best-fit model is shown in \autoref{fig:radio-sed}.

Our second SED was obtained in the VLA A-configuration (the most extended and highest angular resolution; e.g., the L-band synthesized beam is $1.3^{\prime\prime}$), so the host galaxy emission is completely resolved, and therefore, no flux correction is needed. In this epoch, we find $F_{\nu,p}=38.3\pm 2.0$ mJy at $\nu_p=2.36\pm 0.15$\,GHz, with $p=2.67\pm 0.18$ and $\nu_c=14.6^{+3.8}_{-2.9}$\,GHz with a noise parameter of $\ln f=-1.9\pm0.2$. The SED evolution between the two epochs is fairly minor, with mostly an overall decrease in $F_{\nu,p}$; see \autoref{fig:radio-sed}.  This is unsurprising given the relatively small fractional time difference between the two SEDs of $\approx 0.1$.

We note that based on the SED parameters, the expected TDE contribution in the VLITE 340 MHz band is only $\approx 2$ mJy, well below the measured mean value of $\approx 12$ mJy \autoref{sec:host}, thereby supporting our conclusion that the host galaxy dominates the emission in that band.

\subsection{Non-Relativistic Equipartition Analysis \label{subsec:equipartition}}

With the measured SED parameters, we can extract the physical parameters of the outflow by applying an equipartition analysis in the same fashion as \citet[][see Eqns.~$2-6$ therein and Eqns.~27 and 28 in \citealt{barniol_duran_radius_2013}]{cendes_ubiquitous_2024}. In particular, we determine the equipartition energy radius, $R_{\rm eq}$, energy, $E_{\rm eq}$, the magnetic field strength, $B$, and the electron number density of the ambient medium, $n_e=N_{e,{\rm eq}}/4V$, where we assume a spherically symmetric outflow with an emitting shell thickness of $0.1R_{\rm eq}$, implying a surface area filling factor $f_A=1$ and volume filling factor $f_V=\frac{4}{3}(1-0.9^3)\approx 0.36$. We also assume the fraction of post-shock energy in the radiating electrons is $\epsilon_e=0.1$ (typical in analyses of TDE outflow SEDs, e.g., \citealt{cendes_ubiquitous_2024, goodwin_systematic_2025}), with $\epsilon_B=\epsilon_e$ for a measurement assuming equipartition (note that this is different than the minimum energy, which occurs at $\epsilon_B=\frac{6}{11}\epsilon_e$).

We show the posterior distributions of the physical parameters, based on the MCMC fits to the SEDs, in \autoref{fig:radio-sed}. We find that the minimal SED evolution maps to consistent equipartition parameters between the two epochs. In particular, we find $R_{\rm eq}\approx 0.078$~pc, $B\approx0.37$~G, and $n_e\approx 19$~cm$^{-3}$.  The energy exhibits a mild rise from $E_{\rm eq}\approx 10^{49.9}$ to $10^{50.4}$~erg, but the posterior distributions overlap. Using the MIR model start time as the launch date, and the average epoch of our SEDs, the inferred mean outflow velocity is $\beta\approx 0.027$. Lacking radio data for the first $\approx1450$~days, the inferred velocity of an outflow delayed by $\sim 1000$~days increases to $\beta\approx 0.039$, still in the non-relativistic regime.

\begin{deluxetable*}{c c c c c c}[t]
\tablecaption{VLBA observation parameters and derived properties \label{tab:vlbi-properties}}
\tablehead{
Epoch & \colhead{Beam} &  \colhead{R.A.} &  \colhead{Decl.} & \colhead{Gaussian Fit} & \colhead{$\chi^2_\nu$} \\
\colhead{[d]} & \colhead{[$a_{\rm FWHM}\times b_{\rm FWHM}$ @ $\phi$]} & \colhead{} & \colhead{} & \colhead{[$\sigma_x\times\sigma_y$ @ $\theta$]} & \colhead{}}
\startdata
\hline
3180 & $1.86\times0.99$\,mas @ $-9.2\degr$ & \hms{19}{49}{24}{8259983} & \dms{+63}{30}{33}{043125} & $0.111\,{\rm pc}\times0.088\,{\rm pc}$ @ $-76.3\degr\pm0.9\degr$ & 1.11 \\
& & $\pm121\,\mu$as & $\pm142\,\mu$as & $(289\pm5\,\mu{\rm as}\times 230\pm6\,\mu{\rm as})$ \\
3569 & $1.85\times1.08$\,mas @ $-8.5\degr$ & \hms{19}{49}{24}{8260181} & \dms{+63}{30}{33}{042897} & $0.127\,{\rm pc}\times0.088\,{\rm pc}$ @ $-76.2\degr\pm0.5\degr$ & 1.06 \\
& & $\pm121\,\mu$as & $\pm142\,\mu$as & $(330\pm6\,\mu{\rm as}\times 228\pm7\,\mu{\rm as})$ \\
\enddata
\end{deluxetable*}

\subsection{Cooling Break 
\label{subsec:cooling}}

From the SED fitting, we find $\nu_c=12.3^{+1.9}_{-1.6}$\,GHz and $\nu_c=14.6^{+3.8}_{-2.9}$\,GHz. The expected value of $\nu_c$ can also be inferred from the source physical properties if the system age is known using \citep{sari_spectra_1998}:
\begin{equation} \label{eq:nu_c}
    \nu_c \approx 2.25\times10^{5}B^{-3}\Gamma^{-1}t_{d}^{-2}\ {\rm GHz},
\end{equation}
where $\Gamma\approx 1$ is the bulk Lorentz factor and $t_d$ is the time since particle acceleration in days. Using the best-fit parameters, and assuming an outflow launched at the time of disruption, we find an expected value of $\nu_c \approx 0.39$\,GHz using the mean values of the two SED epochs. 

This discrepancy can be reconciled in two possible ways. First, if the outflow is younger (i.e., a delayed launch). In this scenario, using the ratio of predicted to observed values of $\nu_c$ for the two epochs, we find that the outflow would need to be a factor of 5.9 times younger, equivalent to a launch date $\approx 440$~days prior to the first SED observation (i.e., at about $2820$~days). This is clearly in conflict with the first radio detection at 1454~days. Thus, a delayed launch may account for some, but not all, of the discrepancy in the value of $\nu_c$; a delay of $\sim 1000$~d would only increase the predicted value of $\nu_c$ by a factor of about 2.

The second way to reconcile the discrepancy is a deviation from the equipartition requirement of $\epsilon_e=\epsilon_B=0.1$. From \citet{barniol_duran_radius_2013}, we find that $B\propto R_{\rm eq}^4\propto (\epsilon_B/\epsilon_e)^{4/17}$, and therefore, $\nu_c\propto (\epsilon_B/\epsilon_e)^{-12/17}$. Thus, the discrepancy between the observed and predicted values of $\nu_c$ can be reconciled if $\epsilon_B\approx 0.007\epsilon_e$. This is about an order of magnitude lower than inferred in other non-relativistic TDEs \citep[e.g.,][]{cendes_mildly_2022}. Taking into account the deviation from equipartition, the physical parameters for the first epoch are modified to the following values: $R_{\rm eq}\approx 0.062\pm0.003$~pc, $E_{\rm eq}\approx 10^{50.4\pm0.2}$~erg, $B\approx 0.073\pm0.013$~G, and $n_e\approx 245\pm27$~cm$^{-3}$. For the second epoch, the values become the following: $R_{\rm eq}\approx 0.057\pm0.004$~pc, $E_{\rm eq}\approx 10^{51\pm0.3}$~erg, $B\approx 0.17\pm0.06$~G, and $n_e\approx 365\pm50$~cm$^{-3}$. The velocity inferred from the mean radius is correspondingly lower, with a value of $\beta\approx0.021$ for no delay, and $\beta\approx0.030$ for a 1000-day delay.

\subsection{VLBI Analysis 
\label{subsec:vlbi-analysis}}

\begin{figure}[t!]
    \centering
    \includegraphics[width=\linewidth]{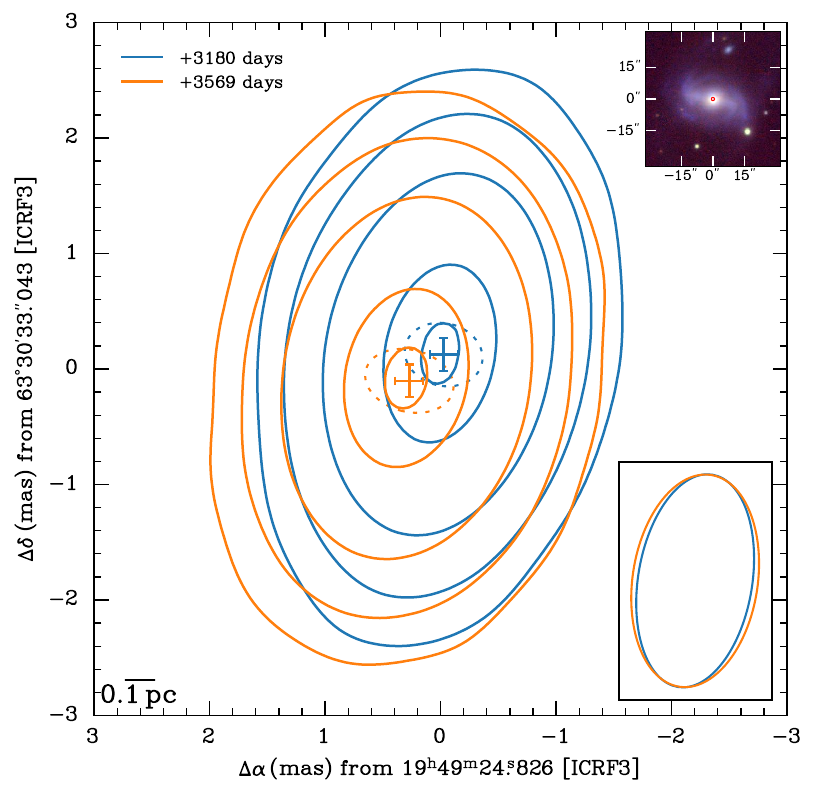}
    \caption{Contour plots of the self-calibrated images reconstructed from our VLBA observations on 2024 January 2 (blue) and 2025 January 25 (orange), where the maps have been shifted to the elliptical Gaussian's position that best fits the pre-self-calibrated data. Contours are displayed at 95, 64, 16, 4, and 1 percent of the peak flux density to allow for a uniform, noise-independent visual comparison.  The synthesized beam of each observation is shown in the bottom-right inset. The best-fitting parameters of the elliptical Gaussian for each epoch are listed in \autoref{tab:vlbi-properties}. In the top-right inset, we also show the location of the radio source on the host galaxy PanSTARRS1 color-stacked image \citep{flewelling_pan-starrs1_2020}, indicating it is located in the galaxy nucleus.}
    \label{fig:vlbi}
\end{figure}

We provide the best-fitting position and size of a single elliptical Gaussian for the two VLBA epochs in  \autoref{tab:vlbi-properties}, and show the self-calibrated contour plots in \autoref{fig:vlbi}. We note that the inferred source sizes from the visibility fitting routine are slightly smaller than the synthesized beam, so our measurements are subject to the assumption of an underlying Gaussian source profile. Future VLBI observations may reveal a more complex morphology as the source continues to expand. We also note that changes to the underlying brightness distribution can result in a degeneracy between a source shift and a change in size when the majority of the brightness distribution is encompassed within a single beam. For example, a dimming of one side of a quasi-spherical outflow (as often seen in VLBI observations of supernovae remnants) at the sub-resolution scale could appear like a Gaussian's position shifting to the new flux-weighted centroid of the extended emission. Therefore, an observation of astrometric motion does not strictly rule out non-relativistic quasi-spherical outflows, but sub-resolved observations can constrain early indicators of actual source motion.

Using the phase referencing systematic astrometric uncertainty tables from \citet{pradel_astrometric_2006}, we can conservatively estimate the astrometric error using the separation on the sky of the primary phase calibrator J2006+6424 to \wtp. These systematic uncertainties dominate over the statistical error from the visibility fit; we list their values in \autoref{tab:vlbi-properties}. The centroid shifts by $\Delta\alpha=+0.30$\,mas, $\Delta\delta=-0.23$\,mas. Using the uncertainties and combining with the measured shift, we find a Mahalanobis distance of $r\approx2.07$, corresponding to a $p$-value of $\approx0.12$ from the Rice distribution with $k=2$; therefore, we cannot yet state whether the measured shift is significant.

The source size along its major axis expands from $\approx0.11$~pc to $\approx0.13$~pc, or an inferred expansion velocity of $\beta\approx0.049$. Although we don't detect expansion along the source's minor axis, this is unsurprising since the major axis of the synthesized beam is more aligned with the source minor axis. The mean VLBI-derived size is nearly a factor of two times larger than the mean value derived from the SED (when including the deviation from equipartition). A mild disagreement is not surprising, given the various assumptions that enter into the equipartition modeling; any deviation from the assumed geometry of a spherical outflow ($f_A=1$, $f_V=0.36$) or a thinner/thicker shock width than the assumed 10\% could modify the characteristic size but still be consistent with the observed SEDs.

\begin{figure}[t!]
    \centering
    \includegraphics[width=\linewidth]{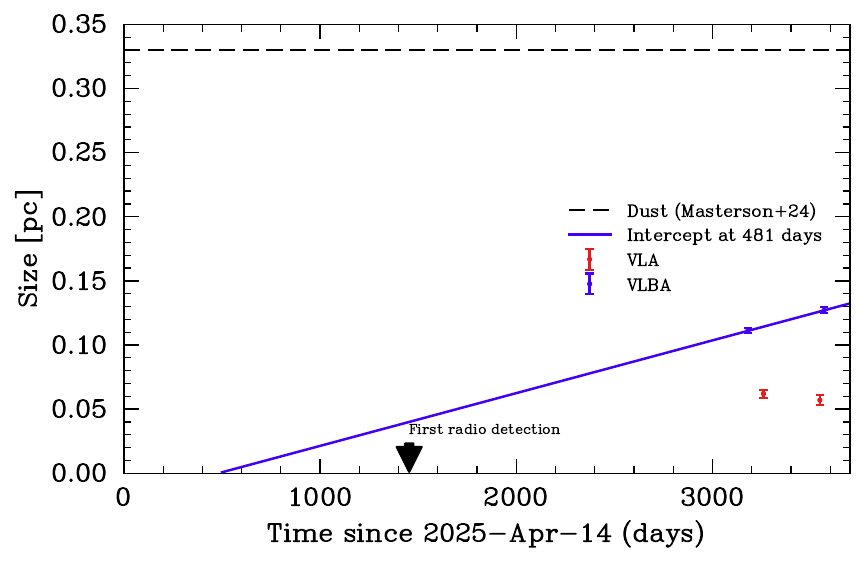}
    \caption{The radius evolution inferred from our VLBA observations (blue) and SED-based equipartition modeling (red). A basic linear fit to the VLBA measurements indicates a launch time of $\sim 500$~days after disruption, roughly a $\sim1000$~days after the first radio detection in VLASS. Also shown is the inferred radius of the dust shell that produces the reprocessed MIR emission \citep{masterson_new_2024}. We caution that the velocity estimate and intercept are highly uncertain, given that only two measured radii are available.}
    \label{fig:radii}
\end{figure}

The expansion rate measured from the two epochs is larger than the mean expansion rate inferred from the size in each epoch and the assumption of a prompt launch at $t=0$. Indeed, the mean expansion rate is $\approx 0.0416c$ and $\approx 0.0424c$ from the 3180\,d and 3569\,d measurements, respectively. This may be indicative of a delayed launch of the outflow. Fitting the two VLBA measurements, we find a nominal delay of $\sim 500$\,d, about 1000\,d before the first radio detection with an average velocity of $\beta\approx0.049$; see \autoref{fig:radii}.  We caution that this is a high-uncertainty estimate given the availability of only two measurements within about a year to determine the velocity. Upcoming planned VLBA observations will provide a more accurate measurement of the delay time and expansion velocity.

\subsection{Off-Axis Jet? \label{subsec:off-axis}}

The significantly delayed peak of the radio light curve ($\sim 6$ years) could possibly be interpreted as the signature of an off-axis jet \citep[e.g.,][]{granot_off-axis_2018,matsumoto_generalized_2023}, launched at the onset of the MIR flare, or with a delay of up to $\sim 1000$\,d (set by the initial radio detection). 

\subsubsection{Prompt Launch}

\citet{granot_off-axis_2018} simulated the light curves and radio images of an off-axis jet at several viewing angles, assuming an isotropic kinetic energy of $E_K=10^{53}$ erg, a typical value inferred from observations of on-axis jetted TDEs \citep[e.g., Sw\,J1644+57][]{cendes_radio_2021}. Even for the extremal off-axis angle of $90^\circ$, the radio light curves are expected to peak on a timescale of $\lesssim 2$ years, shorter than observed for \wtp. The time at which the light curves peak scales with the jet energy as $\propto E^{1/3}$, requiring more than an order of magnitude increase in the jet energy to match the timescale in \wtp; smaller off-axis angles would require an even larger increase in the energy budget--the size of the radio image scales by the same energy correction factor. For a $90^\circ$ off-axis jet of this energy, the predicted size is $\approx 2$ pc \citep{granot_off-axis_2018}, more than an order of magnitude larger than the measured maximum size of $\approx 0.13$ pc.  We can therefore rule out an off-axis jet launched at the time of tidal disruption.

\begin{figure*}[t]
    \centering
    \includegraphics[width=\linewidth]{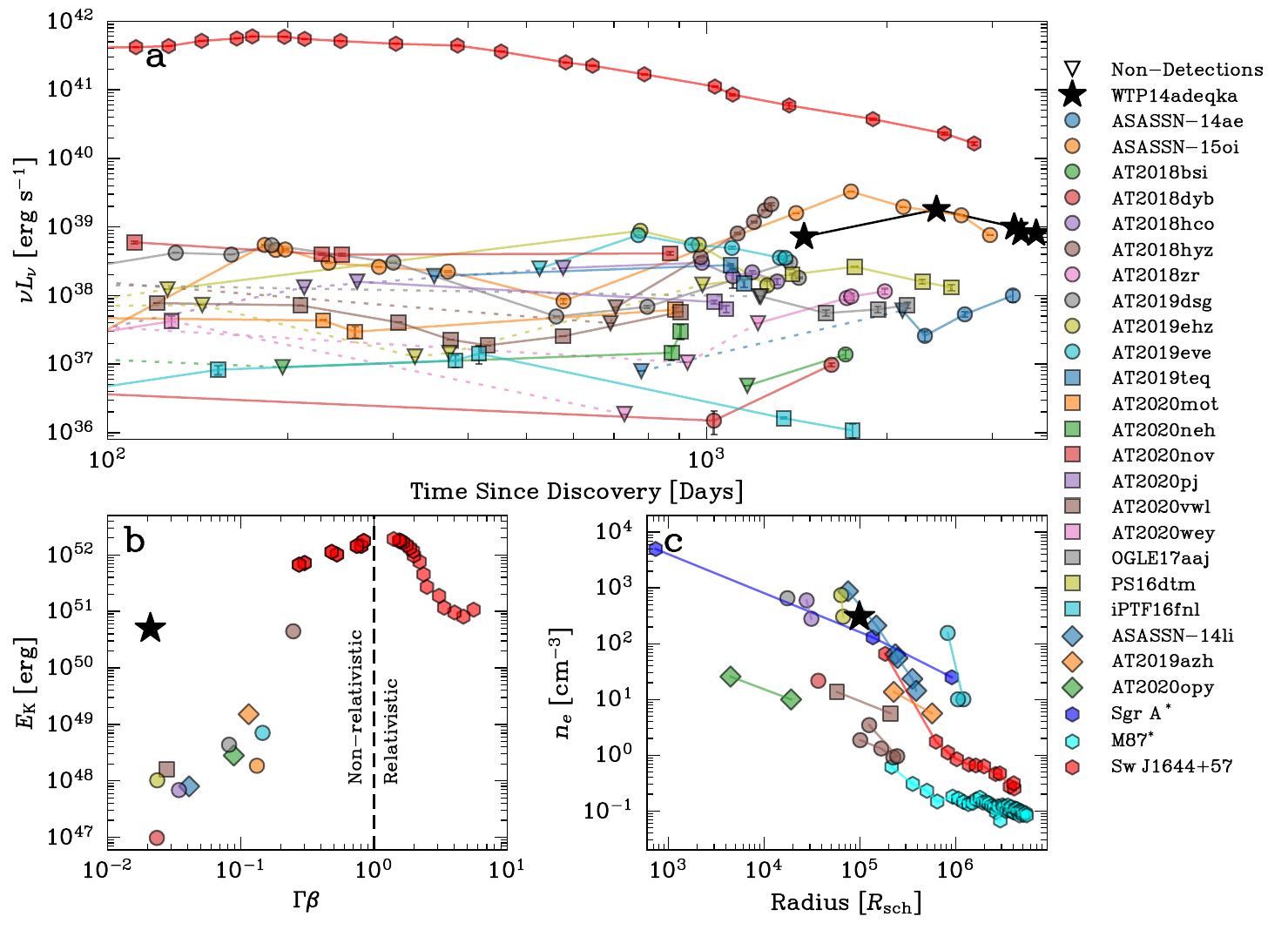}
    \caption{A comparison of the \wtp's radio outflow properties extracted from the SED equipartition analysis relative to the optical sample of \citet[][Figures 1, 6, and 7]{cendes_ubiquitous_2024} with additions to ASASSN-15oi from \citet{hajela_eight_2024}. \wtp\ falls within parameters typical for a non-relativistic TDE. (a) The radio luminosity light curves of optically-discovered TDEs with a delayed radio brightening compared to the S-band luminosity of \wtp. (b) Kinetic energy of the outflow versus the outflow velocity of optical TDEs compared with \wtp. The energy is strikingly larger than other non-relativistic TDEs, despite a low inferred velocity from a launch date coeval with the MIR discovery date. (c) The inferred radial density profiles of the optical TDEs alongside results from the Milky Way and M87. \wtp\ falls squarely in the typical range.}
    \label{fig:comparison}
\end{figure*}

\subsubsection{Delayed Launch}

If we allow the jet to be launched with a delay of $\sim 1000$\,d, we can potentially avoid the requirement of a $\approx 90^\circ$ off-axis angle and enter the regime where the emission is dominated by either the jet or counterjet, leading to a smaller image size. In this scenario, we expect both an astrometric shift in the image centroid and expansion of the image size itself. We consider the \citet{granot_off-axis_2018} model with an off-axis angle of $\theta_{\rm obs}=0.8$ ($45^\circ$), delayed by $\sim 1000$\,d. At the resulting ``corrected'' timescale, the emission is dominated by the counterjet \citep[see Figure~4 of][at 2195\,d]{granot_off-axis_2018}. The counterjet image centroid is expected to exhibit an outward shift to a maximum separation of $\approx 1$\,pc from the launch origin, and then appear to reverse direction back toward the launch origin at later times as the jets become non-relativistic and contribute similarly. The timescale at which the apparent motion is negligible is $\sim 2000-3000$~days, in the regime of our VLBA observation. Thus, it is in principle possible that we are observing the turnover in the image centroid, with a negligible astrometric shift. We would expect in this scenario a shift of $\sim 0.7$~pc (1.8 mas) in the next $\sim 1000$~days, which will be easily measurable in upcoming VLBA observations.

A more significant challenge to the model, however, is that the counterjet's image size exhibits significant expansion with an apparent velocity of $\gtrsim 0.12c$, roughly three times larger than any of our measured values. The expansion rate increases with time, to $\approx 0.3c$ in $\sim 1000$~days from our most recent VLBA epoch, again measurable with planned future VLBA observations.

Observational evidence of other possible off-axis jets in TDEs is similarly incongruent with our data. Arp~299-B~AT1 ($d\approx40$\,Mpc) was constrained to a moderate viewing angle of $25^\circ-35^\circ$ by the lack of a detected counterjet. It was observed to exhibit both the predicted motion and expansion of the radio source at $\gtrsim 7$~years after the MIR transient \citep[][see their Figure~1]{mattila_dust-enshrouded_2018}. The VLBA images at seven and eight years display a clear shift of nearly 1~mas for an average expansion velocity of $\beta_{\rm obs}=0.25$ over the ten years of VLBA observations post-disruption, indicating outward astrometric motion should be detectable in \wtp\ given its current age (\wtp\ is only twice the distance of Arp~299-B~AT1). Additionally, the images at the same age are far more extended (at least a factor of three times larger than our images of similar age), indicating that the source size (even at intermediate off-axis angles) would be much larger than we find. Finally, the decelerating jet of Arp~299-B~AT1 is considerably more asymmetric, with a discernible direction of travel. 

We therefore conclude that an off-axis jet with a delayed launch of $\sim 1000$~days could possibly be accommodated with our VLBA observations with some fine-tuning of model parameters, but upcoming VLBA observations will provide a definitive test.

\section{Comparison to Optical TDE Outflows\label{sec:discussion}}

To contextualize the radio emission and inferred outflow properties of \wtp, in \autoref{fig:radio-sed} we compare our results to the sample of optical TDEs from \citet{cendes_ubiquitous_2024}.  We find that the radio emission from \wtp\ has a peak luminosity ($\approx 2\times 10^{39}$ erg s$^{-1}$) at the upper end of the distribution of optical TDEs (matched only by ASASSN-15oi and AT2018hyz; \citealt{hajela_eight_2024, cendes_ubiquitous_2024}), and a peak timescale ($\approx 6$ years) comparable to the latest-peaking optical TDEs. However, the radio light curve remains an order of magnitude less luminous than the on-axis relativistic TDE Sw\,J1644+57 \citep{eftekhari_radio_2018, cendes_radio_2021} at a comparable timescale.

We also show in \autoref{fig:comparison} the outflow kinetic energy ($E_K$) and velocity ($\Gamma\beta$) in relation to those of optical TDEs and Sw\,J1644+57 \citep{zauderer_illuminating_2013, alexander_discovery_2016,eftekhari_radio_2018,stein_tidal_2021,cendes_radio_2021, cendes_radio_2021-1, cendes_mildly_2022, goodwin_at2019azh_2022, goodwin_radio-emitting_2023,cendes_ubiquitous_2024,hajela_eight_2024}. The outflow in \wtp\ is more energetic by about an order of magnitude, but of comparable velocity to those of optical TDEs; the only exception is AT2018hyz with a higher energy and velocity \citep{cendes_mildly_2022}. 

Finally, we compare the ambient density to those inferred for optical TDEs and Sw\,J1644+57, scaling the radius by the SMBH Schwarzschild radius estimated by \citet{masterson_new_2024} of $\log M_{\rm SMBH}\approx6.8\pm0.5$ (\autoref{fig:comparison}). We find that the density is at the high end of the range inferred for previous TDEs; this is unsurprising given the anticipated dusty environment. Given the slow expansion and small fractional separation of our observations, we cannot yet assess the radial density profile.

\section{Conclusions \label{sec:summary}}

We presented the discovery and detailed study of a long-duration radio transient associated with the MIR TDE \wtp. We studied the source using both multi-frequency VLA observations and high-resolution VLBA observations. Our key findings are summarized as follows:
\begin{itemize}
    \item The radio emission peaks on a timescale of $\sim 5-6$ years, with a luminosity of $\sim 2\times 10^{39}$ erg s$^{-1}$, making it one of the most delayed and luminous radio counterparts to a TDE. The radio emission subsequently fades as $t^{-2}$, with our most recent radio observations at $\approx 10$ years still exhibiting bright emission.
    \item Optical spectroscopy and a comparison of pre-transient radio and FIR detections of the host galaxy rule out pre-existing AGN activity; the pre-existing radio emission is due to spatially-extended star formation, and is resolved out in most of the TDE follow-up observations.
    \item Two epochs of multi-frequency radio observations at 8.9 and 9.7 years reveal a steady peak frequency at $\approx 2$\,GHz, with mild fading. Based on a detection of a cooling break, we find a deviation from equipartition ($\epsilon_B\approx 0.007\epsilon_e$) and the following physical parameters for the outflow: $R_{\rm eq}\approx 0.06$~pc, $E_{\rm eq}\approx 10^{50.7}$, $B\approx 0.12$~G, and $n_e\approx 305$~cm$^{-3}$. The inferred velocity is $\beta\approx0.021$ for no delay, and $\beta\approx0.030$ for a 1000-day delay.
    \item We directly resolve the radio source in two epochs of VLBA observations at 8.7 and 9.8 years, and find a size increase from 0.11 to 0.13 pc, corresponding to a non-relativistic expansion velocity of $\beta\approx0.05$. The size measurements indicate a delayed launch of the outflow by $\approx 2-3$ years.
    \item The relatively small size of the emission region and the lack of astrometric motion rule out emission from a promptly launched, $\sim 90^\circ$ off-axis jet (as would be required for a light curve peak several years after the TDE); such a jet would be an order of magnitude larger.  A delayed off-axis jet with a moderate off-axis angle can be marginally accommodated, but upcoming VLBA observations will be able to directly test this scenario.
    \end{itemize}

Delayed radio emission appears to be common in optical TDEs, and now, for the first time, is also seen in a MIR TDE. Our ability to directly resolve the radio-emitting region, combined with detailed multi-frequency observations, implicates a delayed, non-relativistic outflow as the most likely origin of the emission. Continued monitoring of this source will reveal its dynamical evolution, and radio observations of the full sample of MIR TDEs (Golay et al.,~in prep.) will reveal the ubiquity of delayed radio emission in these events.

\begin{acknowledgments}
The \href{https://public.nrao.edu/}{National Radio Astronomy Observatory} is operated by Associated Universities Inc., under cooperative agreement with the National Science Foundation. This research has made use of NASA's \href{https://ui.adsabs.harvard.edu/}{Astrophysics Data System}. This work made use of the DiFX software correlator developed at Swinburne University of Technology as part of the Australian Major National Research Facilities program \citep{deller_difx-2_2011}. The Berger Time-Domain Group at Harvard is supported by NSF and NASA grants. Basic research in Radio Astronomy at the U.S.\ Naval Research Laboratory is supported by 6.1 Base funding. Construction and installation of VLITE was supported by the NRL Sustainment Restoration and Maintenance fund.
\end{acknowledgments}

\vspace{5mm}
\facilities{Karl G. Jansky Very Large Array (VLA; NRAO), VLA Low-band Ionosphere and Transient Experiment (VLITE; NRAO), Very Long Baseline Array (VLBA; NRAO), Binospec (MMT Observatory)}

\software{\aips\ \citep[version 31DEC2024,][]{greisen_aips_2003}, Astropy \citep[v6.1.7,][]{astropy_collaboration_astropy_2013, astropy_collaboration_astropy_2018, astropy_collaboration_astropy_2022}, Cmcrameri \citep[v1.9,][]{crameri_scientific_2023}, {\tt difmap} \citep[v2.5q,][]{shepherd_difmap_1997}, Emcee \citep[v3.1.6,][]{foreman-mackey_emcee_2013}, Matplotlib \citep[v3.10.0,][]{hunter_matplotlib_2007}, Numpy \citep[v2.2.0,][]{harris_array_2020}, Scipy \citep[v1.14.1,][]{virtanen_scipy_2020}, and Smplotlib \citep[v0.0.9,][]{li_astrojacoblismplotlib_2023}}

\startlongtable
\begin{deluxetable*}{l l c c c c}
\tablecaption{Radio flux densities of \wtp\ reported in this work \label{tab:fluxes}}
\tablehead{\colhead{Telescope} & \colhead{Project Code} & \colhead{Date of Observation} & \colhead{$\delta t$ [days]} & \colhead{Frequency [GHz]} & \colhead{Flux Density [mJy]} }
\startdata
\hline
VLITE & VCSS1.2 & 2019 Apr. 12 & 1454 & 0.34 & $18.6\pm3.4$ \\
--- & VCSS2.2 & 2021 Dec. 01 & 2418 & --- & $15.8\pm3.4$ \\
--- & VCSS3.2 & 2024 Jun. 19 & 3349 & --- & $<14.3$ \\
\hline
VLA & VLASS1.2 & 2019 Apr. 12 & 1454 & 3 & $29.56\pm0.12$ \\
--- & VLASS2.2 & 2021 Dec. 01 & 2418 & --- & $73.68\pm0.13$ \\
--- & VLASS3.2 & 2024 Jun. 19 & 3349 & --- & $33.87\pm0.13$ \\
\hline
VLA & 23B-340 & 2023 Dec. 21 & 3168 & 9 & $22.58\pm0.02$ \\
--- & 23B-340 & 2023 Dec. 21 & 3168 & 11 & $20.16\pm0.028$ \\
\hline
VLITE & 24A-328 & 2024 Mar. 22 & 3260 & 0.34 & $10.8\pm2.6$ \\
VLA & --- & --- & --- & 1.125 & $41.93\pm1.21$ \\
--- & --- & --- & --- & 1.375 & $46.97\pm0.21$ \\
--- & --- & --- & --- & 1.625 & $46.47\pm0.28$ \\
--- & --- & --- & --- & 1.875 & $44.32\pm0.12$ \\
--- & --- & --- & --- & 2.25 & $44.48\pm0.10$ \\
--- & --- & --- & --- & 2.75 & $42.18\pm0.06$ \\
--- & --- & --- & --- & 3.25 & $39.58\pm0.05$ \\
--- & --- & --- & --- & 3.75 & $35.89\pm0.06$ \\
--- & --- & --- & --- & 4.5 & $32.20\pm0.05$ \\
--- & --- & --- & --- & 5.5 & $28.77\pm0.06$ \\
--- & --- & --- & --- & 8.5 & $22.66\pm0.05$ \\
--- & --- & --- & --- & 9.5 & $20.90\pm0.05$ \\
--- & --- & --- & --- & 10.5 & $19.08\pm0.06$ \\
--- & --- & --- & --- & 11.5 & $17.23\pm0.06$ \\
--- & --- & --- & --- & 13 & $15.79\pm0.04$ \\
--- & --- & --- & --- & 15 & $13.74\pm0.04$ \\
--- & --- & --- & --- & 17 & $12.32\pm0.04$ \\
--- & --- & --- & --- & 20 & $10.19\pm0.05$ \\
--- & --- & --- & --- & 24 & $8.19\pm0.05$ \\
--- & --- & --- & --- & 31 & $5.45\pm0.06$ \\
--- & --- & --- & --- & 35 & $4.65\pm0.06$ \\
--- & --- & --- & --- & 42 & $4.10\pm0.12$ \\
--- & --- & --- & --- & 46 & $2.82\pm0.19$ \\
\hline
VLITE & 24B-448 & 2024 Nov. 09 & 3492 & 0.34 & $16.9\pm3.7$ \\
\hline
VLITE & 24B-091 & 2025 Jan. 03 & 3547 & 0.34 & $9.3\pm2.8$ \\
VLA & -- & -- & -- & 1.125 & $25.78\pm0.13$ \\
-- & -- & -- & -- & 1.375 & $30.28\pm0.08$ \\
-- & -- & -- & -- & 1.625 & $33.00\pm0.12$ \\
-- & -- & -- & -- & 1.875 & $34.78\pm0.09$ \\
-- & -- & -- & -- & 2.25 & $33.82\pm0.09$ \\
-- & -- & -- & -- & 2.75 & $33.81\pm0.06$ \\
-- & -- & -- & -- & 3.25 & $32.44\pm0.06$ \\
-- & -- & -- & -- & 3.75 & $30.25\pm0.06$ \\
-- & -- & -- & -- & 4.5 & $28.02\pm0.05$ \\
-- & -- & -- & -- & 5.5 & $25.41\pm0.05$ \\
-- & -- & -- & -- & 7.5 & $22.21\pm0.05$ \\
-- & -- & -- & -- & 8.5 & $20.03\pm0.06$ \\
-- & -- & -- & -- & 9.5 & $18.55\pm0.05$ \\
-- & -- & -- & -- & 10.5 & $16.83\pm0.05$ \\
-- & -- & -- & -- & 11.5 & $15.34\pm0.07$ \\
-- & -- & -- & -- & 13 & $14.68\pm0.04$ \\
-- & -- & -- & -- & 15 & $12.54\pm0.04$ \\
-- & -- & -- & -- & 17 & $10.08\pm0.06$ \\
-- & -- & -- & -- & 20 & $9.13\pm0.04$ \\
-- & -- & -- & -- & 24 & $7.52\pm0.04$ \\
-- & -- & -- & -- & 31 & $3.90\pm0.05$ \\
-- & -- & -- & -- & 35 & $3.11\pm0.05$ \\
-- & -- & -- & -- & 42 & $1.87\pm0.12$ \\
-- & -- & -- & -- & 46 & $2.16\pm0.19$ \\
\enddata
\tablecomments{Uncertainties are derived from the background r.m.s. of each image and are a better representation of the detection significance. We do not include the typical 7\% absolute flux density scaling uncertainty for VLA observations in this table.}
\end{deluxetable*}

\bibliography{references}{}
\bibliographystyle{aasjournal}

\end{document}